\documentclass[11pt]{article}
\usepackage{ae,lmodern}
\usepackage[utf8]{inputenc} 
\usepackage[T1]{fontenc}

\usepackage{geometry}
\usepackage{xcolor}
\usepackage{authblk}
\usepackage{amsmath}
\usepackage{amssymb}
\usepackage{amsthm}
\usepackage{graphicx}
\usepackage{caption}
\usepackage{comment}
\usepackage{accents} 
\usepackage{float}
\usepackage{lineno}
\usepackage{setspace}
\usepackage[authoryear]{natbib}
\usepackage[colorlinks=true, allcolors=blue]{hyperref}
\usepackage{cleveref}
\newcommand{\ie}{i.e., }
\newcommand{\eg}{e.g., }
\newcommand{\mP}{\mathbb{P}}

\title{Independence in Integrated Population Models}

\author{Frédéric Barraquand}
\affil{\normalsize Institute of Mathematics of Bordeaux,\\
University of Bordeaux, CNRS, Bordeaux INP, Talence, France}
\date{\vspace{-5ex}}

\begin{document}
\maketitle
\thispagestyle{empty}
\begin{abstract}
 Integrated population models (IPMs) combine multiple ecological data types such as capture-mark-recapture histories, reproduction surveys, and population counts into a single statistical framework. In such models, each data type is generated by a probabilistic submodel, and an assumption of independence between the different data types is usually made. The fact that the same biological individuals can contribute to multiple data types has been perceived as affecting their independence, and several studies have even investigated IPM robustness in this scenario. However, what matters from a statistical perspective is probabilistic independence: the joint probability of observing all data is equal to the product of the likelihoods of the various datasets. Contrary to a widespread perception, probabilistic non-independence does not automatically result from collecting data on the same physical individuals. Conversely, while there can be good reasons for non-independence of IPM submodels arising from sharing of individuals between data types, these relations do not seem to be included in IPMs whose robustness is being investigated. Furthermore, conditional rather than true independence is sometimes assumed. In this conceptual paper, I survey the various independence concepts used in IPMs, try to make sense of them by getting back to first principles in toy models, and show that it is possible to obtain probabilistic independence (or near-independence) despite two or three data types collected on the same set of biological individuals. I then revisit recommendations pertaining to component data collection and IPM robustness checks, and provide some suggestions to bridge the current gap between individual-level IPMs and their population-level approximations using composite likelihoods. 
\end{abstract}

\textbf{Keywords}: Integrated Population Model; independence; data fusion; reproduction; survival, Cormack-Jolly-Seber. \\ 


\section{Introduction}
An integrated population model gathers several data streams into a single statistical framework for demographic parameter estimation. Typically, within population ecology at least, the datasets considered are counts of individual organisms (\eg a population census), capture-mark-recapture or mark-recovery data, and sometimes fecundity data (\eg a survey of nests). The likelihoods corresponding to each dataset are then usually multiplied to provide a single likelihood for the integrated population model, which is then maximized to find the parameter values; or the code for the various models is written sequentially inside a procedure for Bayesian Monte Carlo estimation, with similar assumptions and results. This statistical model construction assumes that it is sensible to multiply component likelihoods in the first place, which can be justified by an assumption of independence between datasets. In their seminal paper on integrated population models, \citet{besbeas2002integrating} made the remark that \textit{``The assumption of independence made here is not likely to be violated''}. Later they more cautiously pointed out that \textit{``It is sometimes the case that census data and demographic data are not independent. This can occur, for instance, in studies involving mammals living on small islands [Soay sheep on St-Kilda, \citet{tavecchia2009estimating}]''} in \citet{besbeas2009completing}. The small study area implies that data on specific sheep individuals can be collected in the form of capture and reproduction histories, while the same individuals are also being counted every year. They found that combining recovery information with dependent [their wording] census data, then \textit{``RMSEs are mostly increased compared to the case of combining with independent census data.''} Dependence referred in that case to the overlap in the sets of individuals that contribute to each dataset. This dependence-through-overlapping-individuals could appear even more worrying when combining capture-mark-recapture and fecundity data, e.g., in birds, since ecologists usually ring and measure fecundity at the nest on the very same individuals. Thus there is potential for massively overlapping sets of individuals contributing to the different datasets. Numerous IPM studies---original work and reviews---have since then equated sharing of individuals between datasets with the independence assumption, and logically warned that the assumption might not be met \citep[e.g.,][]{schaub2011integrated,gamelon2016density,zipkin2018synthesizing,gould2024unified}. 

Due to this perceived violation of IPM assumptions by most available ecological data, simulation-based studies have attempted to check the extent of the issues generated by the sharing of individuals between the datasets on which IPMs are built. They found overall a rather marked robustness. \citet{abadi2010assessment} noted that \textit{``the violation of the assumption of independence had only minor consequences on the precision and accuracy of the parameter estimates''}. Attempts to compare exact individual-based modelling of the likelihood to multiplication of component likelihoods by \citet{beliveau2016data} also yielded similar results, with only small effects on parameter precision.
More recently, \citet{weegman2021partial} stated that \textit{``Contrary to previous investigators, we found no substantive bias or uncertainty in any demographic rate from IPMs derived from data sets with complete overlap [of individuals, between data sources].''} They concluded that the violation of the independence assumption should not dissuade researchers from the application of IPMs in ecological research---see also \citealt{schaub2021integrated}, chap. 7 p. 287, for a detailed summary of the literature on robustness checks of the independence assumption.

In this paper, I attempt to make sense of these apparently contradictory findings and disambiguate the various meanings of the word independence in statistical ecology. I start by asking whether the statistical independence assumption can indeed be violated by simply having the same individuals contributing to different datasets. First in really simple toy models, then slightly more complex ones, and eventually in a very basic integrated population model. I then outline practical recommendations and suggestions for further statistical/theoretical work. 

\section{Probabilistic independence in simple models}

\subsection{Definitions}

Let us start by a couple of definitions of independence concepts. \textbf{Probabilistic or statistical independence} between two events $A$ and $B$ is equivalent to $\mathbb{P}(A \cap B) = \mathbb{P}(A) \mathbb{P}(B)$. The concept is easily extended to any pair of random variables, \eg $\mathbb{P}(X=1, Y=6) = \mathbb{P}(X=1) \mathbb{P}(Y=6)$ for two discrete-valued independent random variables $X$ and $Y$. This can be extended to absolutely continuous random variables $X$ and $Y$ using probability densities, \ie $f_{X,Y}(x,y) = f_X(x) f_Y(y)$. 

The above concept can be constrasted from \textbf{physical independence}, a concept constructed for the occasion of this paper, by which we mean that the physical entities on which the biometric measurements or counts are performed are not the same in the various datasets. We shall also assume that dependence is the logical complement of independence for the two abovementioned concepts. 

We consider a number of datasets $\mathbf{D}_k$, $k \in \{1,...,K\}$ (typically $K=2$ to $4$), each $\mathbf{D}_k$ with values taken in $\mathbb{N}^{d_k} \subset \mathbb{R}^{d_k}$, with $d_k = \dim(\mathbf{D}_k)$, and produced by a model $\mathcal{M}_k$. Each model $\mathcal{M}_k$ has an associated parameter set $\boldsymbol{\theta}_k \in \mathbb{R}^{m_k}$, $m_k$ being the number of parameters in the $k^{\text{th}}$ submodel, possibly with $\boldsymbol{\theta}_k \cap \boldsymbol{\theta}_l \neq \emptyset$ for $k \neq l$. An Integrated Population Model (IPM) is constituted of the various submodels $\mathcal{M}_k$, has parameter set $\boldsymbol{\theta} = \boldsymbol{\theta}_{1} \cup \boldsymbol{\theta}_{2} \cup ... \cup \boldsymbol{\theta}_{K}$ and can be specified at the individual level with the \textbf{full likelihood} $\mathcal{L}(\boldsymbol{\theta};\mathbf{D}_1,...,\mathbf{D}_K) = \mathbb{P}(\mathbf{D}_1,...,\mathbf{D}_K|\boldsymbol{\theta})$, with a slight abuse of notation as I omit for simplicity the random variables generating the observed datasets. This likelihood is a particular case of joint likelihood in the terminology of \citealt{frost2023integrated}. This full likelihood could be very complex if defined through an individual-level stochastic process, and fitted through MCMC. However, to keep a tractable and computationally manageable IPM, even when MCMC is used, most often an IPM is defined and fitted to data using a \textbf{composite likelihood} or \textbf{component likelihood product} which is a product of the likelihoods for the various datasets,
\begin{equation}\label{eq:composite-likelihood}
\mathcal{L}(\boldsymbol{\theta}_1;\mathbf{D}_1) \times \mathcal{L}(\boldsymbol{\theta}_2;\mathbf{D}_2) \times ... \times \mathcal{L}(\boldsymbol{\theta}_K;\mathbf{D}_K) =
\mathbb{P}(\mathbf{D}_1|\boldsymbol{\theta}_1) \times 
\mathbb{P}(\mathbf{D}_2|\boldsymbol{\theta}_2) \times ... \times \mathbb{P}(\mathbf{D}_K|\boldsymbol{\theta}_K), 
\end{equation}
using an assumption of independence (which should here mean probabilistic independence). The crux of the matter is then that the various datasets $\mathbf{D}_k$ may have been collected on the same individuals, so there is physical dependence. This could be problematic if physical dependence implies probabilistic dependence.  

\subsection{Physical dependence does not imply probabilistic dependence}

The simplest and perhaps more important point that I would like to make here is that physical dependence does not necessarily cause probabilistic dependence. If I play dice today, with a 6 as the only winning outcome, and then I play a fair coin toss tomorrow, it will likely be consensual that the probability of me winning both of these games is $\frac{1}{6} \times \frac{1}{2}$. There is probabilistic independence even though there is complete physical dependence in terms of the biological individuals generating the events whose probabilities are evaluated.

The reader may object that the situation is quite far away from the ones studied in population ecology and would be correct. Let us get there progressively, by first considering that I draw a Poisson random variable $Y \sim  \mathcal{P}(\lambda)$, $\lambda>0$ on day 1, and then I draw on day 2 a Bernoulli random variable $X \sim \mathcal{B}(s)$ with $s \in [0,1]$. The joint probability is then 
\begin{equation}\label{eq:indep_events}
    \mP(Y=k, X=1) = \frac{e^{_\lambda} \lambda^k}{k!} \times s
\end{equation}
even though the same individual made both of these draws of random variables. 

Now the reader may object again, it is too simple: there is only one individual. Let us now assume that there are $N$ individuals, \eg $N$ ducks. We draw a Poisson variable $Y_i \sim \mathcal{P}(\lambda_i)$ for net fecundity and then a Bernoulli variable $X_i \sim \mathcal{B}(s_i)$ for survival status of each (female) individual $i \in \{1,...,N\}$, where $s_i$ is the survival probability. We obtain for the joint probability of all these events
\begin{equation}\label{eq:indep_individuals}
    \prod_{i=1}^{N} \mP(Y_i=k_i, X_i=1) = \prod_{i=1}^{N} \left(\frac{e^{\lambda_i} \lambda_i^{k_i}}{{k_i}!} \times s_i \right) =  \left( \prod_{i=1}^{N} \frac{e^{\lambda_i} \lambda_i^{k_i}}{{k_i}!} \right) \times \left(\prod_{i=1}^{N} s_i \right).
\end{equation}
For notational simplicity we have considered the probability that all mothers survive, more general formulations follow in the next section. Note that although this formula relies on probabilistic independence, there are two kinds of probabilistic independence: first, independence between survival and reproduction events for the same individual, which was already underlying eq.~\ref{eq:indep_events}, and then independence between events occurring to different individuals, which allows to use the outer product. In other words, the joint probability of observing survival and reproduction to a certain level of individuals 1 and 2 is exactly the product of the probabilities of each individual's fate observed separately. As a corollary, individuals 1 and 2 are not more or less likely to get the same outcomes than \eg 1 and 3, and this assumption of homogeneity in the population will be discussed later on.

The formula of eq.~\ref{eq:indep_individuals} makes clear that the joint probability of survival and reproduction for all individuals in a given year is $\mP(\text{reproduction dataset})\mP(\text{survival dataset})$. Denoting $\mathcal{M}_1$ the reproduction model and $\mathcal{M}_2$ the survival model, the full likelihood is equal to a product of the dataset component likelihoods $\mP(\mathbf{D}_1|\boldsymbol{\theta}_1,\mathcal{M}_1)\mP(\mathbf{D}_2|\boldsymbol{\theta}_2,\mathcal{M}_2)$. Thus we obtain probabilistic independence of $\mathbf{D}_1$ and $\mathbf{D}_2$ in spite of complete physical dependence between datasets $\mathbf{D}_1$ and $\mathbf{D}_2$. 

While the above arguments may convince the reader that one can obtain probabilistic independence in spite of complete physical dependence in very simple reproduction-survival models, the situation in population ecology is a little more complex as individuals are not only followed over the years, but also imperfectly detected, which is why we use capture-mark-recapture models in the first place  (capture-recapture, for short). We will therefore try to extend the above reasoning to imperfect detection of individuals monitored over multiple years in the following section. 

\section{Probabilistic independence in a basic IPM}

We will consider two models with imperfect detection of individuals (1) a model with only survival and reproduction, and (2) a model with survival, reproduction, and counts. In each case, the detection/survival submodel is a classical Cormack-Jolly-Seber model (hereafter CJS, see references in \citealt{lebreton1992modeling} for a historical perspective and chap. 7 in \citealt{kery2011bayesian} for an introduction). 


\subsection{Capture-recapture/survival and reproduction}

The random variable $X_{i,t}$, taking values in $\{0,1\}$ (non-detected, detected), is now our \textit{observed} individual state over $T$ years, obeying
    \begin{enumerate}
        \item $X_{i,t}|Z_{i,t} \sim \mathcal{B}(Z_{i,t} p_{i,t})$
        \item $Z_{i,t}|Z_{i,t-1} \sim \mathcal{B}(Z_{i,t-1} \phi_{i,t-1})$
    \end{enumerate}
where $Z_{i,t}$ is the latent individual state (dead or alive) taking values in $\{0,1\}$, $p_{i,t}$ the detection probability at time $t$ and for individual $i$, and $\phi_{i,t}$ the survival probability to the next time step. The CJS model has two layers, and is defined here as a hidden Markov model \citep{mcclintock2020uncovering}. At each time step we have a Bernoulli draw for individual survival and a Bernoulli draw for capture, conditional to the survival state. A recorded capture history $x$ will resemble \eg $x = (10001000100)$ for an individual. We denote the observed history for individual $i$ as $x_i = (x_{i,t})_{t \in \{1,...,T\}}$. 

On top of this, we consider Poisson fecundities $Y_{i,t} \sim \mathcal{P}(\lambda_i)$ over $T$ years for each individual $i$. Such reproduction produces new individuals which we do not follow (for simplicity). 
The full likelihood for reproduction and capture-recapture now becomes 
\begin{equation}\label{surv-repro-indiv-LL}
\begin{split} 
     \underbrace{\prod_{i=1}^{N}}_{\text{(C)}}\Bigg( \; \underbrace{\sum_{(z_{i,1},...z_{i,T}) \in \{0,1\}^T}}_{\text{(B)}} \; \underbrace{\prod_{t=2}^{T} \bigg( \mP(X_{i,t} = x_{i,t}|Z_{i,t} = z_{i,t}) \times}_{\text{(A)}} \\
     \underbrace{\mP(Z_{i,t} = z_{i,t}|Z_{i,t-1} = z_{i,t-1}) \times \mP(Y_{i,t} = y_{i,t}|X_{i,t} = x_{i,t}) \bigg)}_{\text{(A)}}\\ \mP(Z_{i,1} = z_{i,1}) \mP(Y_{i,1} = y_{i,1}|X_{i,1} = x_{i,1}) \Bigg)
\end{split}
\end{equation}
and contains 
\begin{itemize}
    \item (A) the joint probability of the capture-recapture and reproduction histories for a given individual. It includes the previous multiplication of $X$ and $Y|X$ probabilities for each individual. Note that a reproduction measurement $Y_{i,t}$ is conditional to an observed individual state $X_{i,t}$ rather than the latent state $Z_{i,t}$ (because if the individual is not observed in a given year, we cannot observe its fecundity, although it may have reproduced, so $\mP(Y_{i,t} = y_{i,t}|X_{i,t} = 0) = 0$).\footnote{It may be possible to include a latent fecundity state but would require an additional hidden process.}
    \item (B) A sum over all possible latent dead/alive states $z_{i,t}$.
    \item (C) The product of probabilities over all $N$ individuals.
\end{itemize}

One can then take the fecundity data out of the individual histories using commutativity:
\begin{equation}
\begin{split} 
     \prod_{i=1}^{N}\Bigg( \; \underbrace{\sum_{(z_{i,1},...z_{i,T}) \in \{0,1\}^T} \; \prod_{t=2}^{T} \bigg( \mP(X_{i,t} = x_{i,t}|Z_{i,t} = z_{i,t})
\mP(Z_{i,t} = z_{i,t}|Z_{i,t-1} = z_{i,t-1}) \bigg) \mP(Z_{i,1} = z_{i,1})}_{\text{capture history for individual } i}\\ \times \underbrace{\prod_{t=1}^{T} \mP(Y_{i,t} = y_{i,t}|X_{i,t} = x_{i,t}) }_{\text{reproduction history for }i} \Bigg) \\
     = \prod_{i=1}^{N}\Bigg( \mathcal{L}_{\text{capture},i}(\boldsymbol{\varphi}_i;\mathbf{x}_i) \times \mathcal{L}_{\text{repro},i}(\boldsymbol{\lambda}_i;\mathbf{y}_i|\mathbf{x}_i)   \Bigg)\\
     = \bigg( \prod_{i=1}^{N} \mathcal{L}_{\text{capture},i}(\boldsymbol{\varphi}_i;\mathbf{x})\bigg) \times \bigg( \prod_{i=1}^{N} \mathcal{L}_{\text{repro},i}(\boldsymbol{\lambda}_i;\mathbf{y}_i|\mathbf{x}_i) \bigg)\\
      =  \mathcal{L}_{\text{capture}}(\boldsymbol{\varphi};\mathbf{x}) \times \mathcal{L}_{\text{repro}}(\boldsymbol{\lambda};\mathbf{y}|\mathbf{x}). 
\end{split}
\end{equation}
\textit{Thus all capture-recapture and reproduction measurements are performed on the exact same individuals and yet the joint likelihood writes as the product of the likelihoods for both kinds of data}.$\qed$

In the Appendix, we provide a complementary model formulation which does not involve the summation over the latent individual states. In the latter ``marginalized'' formulation, obtaining a full likelihood that can be written as the product of the capture-recapture and reproduction data likelihoods is less obvious, though examples for small $T$ suggest it to be possible.

\subsection{Capture-recapture/survival, reproduction and counts}

Let us define the indicator random variable $W_{i,t} = 1$ if individual $i$ is counted in year $t$. For unmarked individuals, $W_{i,t}|Z_{i,t}=1 \sim \mathcal{B}(\delta)$ i.i.d, with $\delta$ the probability to be counted. We consider a larger population $N_t>P_t$ than the previously $N$ marked individuals (now denoted $N_m$), with $P_t$ the number of marked individuals in the set of live individuals $N_t$. 
We have a remainder $O_t=N_t-P_t$ which corresponds to the other, unmarked live individuals at time $t$. We can multiply the likelihood product of the preceding section by $\mP$(counting unmarked individuals) without making any additional assumption. The individuals that are marked are by definition also ``counted'', though differently, since we can tally up their numbers at any time from the capture histories. We obtain
\begin{equation}\label{eq:big-product-with-counts}
\begin{split} 
  \prod_{i=1}^{N_m}\Bigg( \; \underbrace{\sum_{(z_{i,1},...z_{i,T}) \in \{0,1\}^T} \; \prod_{t=2}^{T} \bigg( \mP(X_{i,t} = x_{i,t}|Z_{i,t} = z_{i,t}) \mP(Z_{i,t} = z_{i,t}|Z_{i,t-1} = z_{i,t-1}) \bigg) \mP(Z_{i,1} = z_{i,1})}_{\text{capture-recapture for individual } i} \\
  \times \prod_{t=1}^{T} \underbrace{\mP(Y_{i,t} = y_{i,t}| X_{i,t} = x_{i,t})}_{\text{reproduction of } i} \underbrace{\mP(W_{i,t} = w_{i,t}|X_{i,t}=x_{i,t})}_{\text{count of marked } i} \Bigg) \times  \underbrace{\prod_{t=1}^T  \prod_{i'=1}^{O_t} \delta^{w_{i',t}} (1-\delta)^{1-w_{i',t}}}_{\text{counts of unmarked individuals}} \\
     = \prod_{i=1}^{N_m}\Bigg(\mathcal{L}_{\text{capture},i}(\boldsymbol{\varphi}_i;\mathbf{x}_i) \times \mathcal{L}_{\text{repro and count},i}(\boldsymbol{\lambda}_i;\mathbf{y}_i|\mathbf{x}_i) \Bigg) \times \mathcal{L}_{\text{count, unmarked}}(\delta;\mathbf{w})\\
     = \bigg( \prod_{i=1}^{N_m} \mathcal{L}_{\text{capture},i}(\boldsymbol{\varphi}_i;\mathbf{x})\bigg) \times \bigg( \prod_{i=1}^{N_m} \mathcal{L}_{\text{repro and count},i}(\boldsymbol{\lambda}_i;\mathbf{y}_i|\mathbf{x}_i) \bigg) \times \mathcal{L}_{\text{count, unmarked}}(\delta;\mathbf{w})\\
      = \mathcal{L}_{\text{capture}}(\boldsymbol{\varphi};\mathbf{x}) \times  \mathcal{L}_{\text{repro}}(\boldsymbol{\lambda};\mathbf{y}|\mathbf{x}) \times
      \prod_{t=1}^T \prod_{i=1}^{O_t} \delta^{w_{i,t}} (1-\delta)^{1-w_{i,t}} 
\end{split}
\end{equation}

The probability of counting the marked individuals $\prod_{i=1}^{N_m} \prod_{t=1}^{T} \mP(W_{i,t}=w_{i,t}|X_{i,t} = x_{i,t})$ is equal to 1, as for any given individual and time $\mP(W=0|X=0) = 1$ (unobserved marked individuals are not counted) and $\mP(W=1|X=1) = 1$ (recaptured marked individuals are counted), and for marked individuals, $W=X$. We now use $W_{i,t}$ only for unmarked individuals. This eventually leads to the above third term in eq.~\ref{eq:big-product-with-counts} when including unmarked individuals, so that
\begin{equation}
\begin{split} 
\mathcal{L}_{\text{full}}
      = \mathcal{L}_{\text{capture}}(\boldsymbol{\varphi};\mathbf{x}) \times \mathcal{L}_{\text{repro}}(\boldsymbol{\lambda};\mathbf{y}|\mathbf{x}) \times
      \prod_{t=1}^T \prod_{i=1}^{O_t} \delta^{w_{i,t}} (1-\delta)^{1-w_{i,t}}\\
       = \mathcal{L}_{\text{capture}}(\boldsymbol{\varphi};\mathbf{x}) \times  \mathcal{L}_{\text{repro}}(\boldsymbol{\lambda};\mathbf{y}|\mathbf{x}) \times
      \prod_{t=1}^T \delta^{\sum_{i=1}^{O_t}w_{i,t}} (1-\delta)^{O_t-\sum_{i=1}^{O_t}w_{i,t}}\\
      =  \mathcal{L}_{\text{capture}}(\boldsymbol{\varphi};\mathbf{x}) \times  \mathcal{L}_{\text{repro}}(\boldsymbol{\lambda};\mathbf{y}|\mathbf{x}) \times
      \mathcal{L}_{\text{count}}(\delta,\mathbf{O};\mathbf{w})
\end{split}
\end{equation}
where $\mathcal{L}_{\text{count}}(\delta,\mathbf{O};\mathbf{w})$ can also be written $\mathcal{L}_{\text{count}}(\delta,\mathbf{O};\mathbf{S})$ with $\mathbf{S} = (S_1,...,S_T)$ and $S_t = \sum_{i=1}^{O_t}w_{i,t}$ the observed unmarked population size. See \citet{lee2015integrated} for a similar decomposition of counts of unmarked vs marked individuals. A different formula could have been used if the counting method also included marked individuals by default (\eg those are counted by plane and we cannot see their marks), with  $\mathcal{L}_{\text{count}}(\delta,\mathbf{N};\mathbf{w})$ where  $\mathbf{N} = (N_1,...N_T)$ is the total population size. In the present setup, true unmarked population size $O_t$ is unobserved and a parameter to estimate. The formulation, which is close to in $N$-mixture models, suggests that $\delta$ should be fixed or strongly informed to estimate $\mathbf{O} = (O_1,...O_T)$ (see \eg chap. 12 in \citealt{kery2011bayesian} for an introduction to $N$-mixture models, \citealt{barker2018reliability} for a discussion). A naïve estimate of the total population size can then be obtained by summing the estimates for the unmarked compartment and the marked compartment of the population, \eg $\hat{N}_t =\hat{O}_t + \frac{1}{\hat{p}_t}\sum_{i=1}^{N_m}x_{i,t}$, with $\hat{p}_t$ the detection probability, the right term in the sum being the number of observed marked individuals corrected by probability of detection. 

We can remark that for the last term in eq.~\ref{eq:big-product-with-counts}, there is proportionality to a Binomial likelihood for the count of $S_t$ individuals among $O_t$ live unmarked individuals. A Binomial likelihood would also have a binomial coefficient ${O_t}\choose{S_t}$. In other words, $\mathcal{L}_{\text{composite}} = {O_t \choose S_t} \times  \mathcal{L}_{\text{full}}$. One could imagine that counting unmarked individuals leads to counting with replacement (since without marks many individuals may not be distinguishable from each other ), in this case, the binomial coefficient should be included within eq.~\ref{eq:big-product-with-counts} in the full likelihood and we get exactly $\mathcal{L}_{\text{composite}} = \mathcal{L}_{\text{full}}$.   

The model above includes counts and therefore estimates of population size but does not include \eg a matrix population model producing the numbers of individuals to be counted. Of course, there is still a demographic process in our individual-based model since we model survival and reproduction, but we model the demography of the marked individuals, not the unmarked ones. Checking whether combining an individual-based capture-recapture and fecundity model, for marked and unmarked individuals alike, would produce a similar match between full and composite likelihood would be a reasonable direction in which to expand this work
. The model with population dynamics requires including a Jolly-Seber (JS) model (\citealt{jolly1965explicit,seber1965note}, chap. 10 in \citealt{kery2011bayesian}) as the baseline capture-recapture model, since the JS also models the unmarked individuals, as well as a latent branching process for the population dynamics. Another direction taken by \citet{beliveau2016data} consists in specifying the model with subgroups (marked in the $t^{th}$ year, unmarked) which allows a more straightforward connection of the full likelihood to the counting data---while including also a JS model of the capture-recapture data. 

\section{Discussion}

We have shown here that probabilistic (or statistical) independence of datasets is not equivalent to physical independence of demographic datasets, by which we mean that datasets are gathered on different biological individuals. The full individual-based likelihood of a simple survival-reproduction model, or that of a slightly more complex capture-recapture-reproduction model, can be expressed as a product of the component likelihood for the survival/capture-recapture data and the component likelihood for the reproduction data, collected on the very same individuals. Thus, if we follow individual survival under imperfect detection and additionally follow the reproduction of marked individuals, there can be perfect statistical independence of datasets despite complete overlap of individuals in both capture-recapture and reproduction datasets. 

This very basic Integrated Population Model (IPM) can be extended to also include counts, and there is then proportionality rather than equality between the full and composite likelihoods, though equality is possible with a slight change in assumptions. These results tend to explain why previous IPMs testing the robustness to having the same individuals in all data types found a marked robustness \citep{abadi2010assessment,weegman2021partial}. Their full likelihood was likely very close to the composite likelihood obtained by multiplying the likelihoods of the various submodels. Therefore, it does not seem to make sense to require by default different individuals contributing to each dataset, in order to fit an IPM specified through a composite likelihood made of a product of component likelihoods. Physical dependence does not imply probabilistic dependence ($P \nRightarrow Q$, with $P$ representing physical dependence and $Q$ probabilistic dependence), and thus physical independence is not necessary for probabilistic independence ($\neg Q \nRightarrow \neg P$). 

Even though physical independence is not a necessary condition to obtain probabilistic independence, it is at times a sufficient one, i.e. physical independence between the sets of individuals contributing to the component datasets is enough to generate probabilistic independence ($\neg P \Rightarrow \neg Q$). If $\mathbf{D}_1$, $\mathbf{D}_2$, and $\mathbf{D}_3$ are datasets collected on different subpopulations of individuals, it is correct that the full likelihood could be written as $\mathbb{P}(\mathbf{D}_1|\boldsymbol{\theta}_1) \mathbb{P}(\mathbf{D}_2|\boldsymbol{\theta}_2) \mathbb{P}(\mathbf{D}_3|\boldsymbol{\theta}_3)$ for the simple models presented here. In fact we have used this property when separating counts of marked vs unmarked individuals earlier. Thus if one wants to be extra careful, it stands to reason that one could require physical independence of datasets when building IPMs. Although this reasoning may have been the original logic behind the attention to the overlap in biological individuals contributing to different datasets, it seems to have generated confusion in the literature, with many researchers apparently equating a sufficient condition to a necessary one. Also, one should be aware that no sharing of individuals between datasets may not always be a sufficient condition for probabilistic independence. If physical independence implies probabilistic independence ($\neg P \Rightarrow \neg Q$), the contrapositive, probabilistic dependence implies physical dependence ($Q \Rightarrow P$), must be true. The latter assertion sounds unlikely to be met in practice, outside of a simulated population: there are plenty of ways to induce probabilistic non-independence without any sharing of individuals. Basically any unmeasured confounding variable affecting multiple datasets could induce some statistical non-independence. Furthermore, datasets coming each from a different subpopulation risk violating another assumption of IPMs, that of shared demography (i.e., equality of demographic parameters) between datasets \citep{schaub2021integrated}.  All things considered, it looks unreasonable to require physical independence of datasets to fit a composite likelihood model, because this aspect of data design, which greatly risks reducing sample sizes for the same research effort, will help meeting one assumption of the model, but will do so under restrictive assumptions, while creating a risk to violate another key assumption of IPMs, shared demographic parameters among submodels. Thus it may be wiser to view probabilistic and physical dependence of datasets as two different concepts, although there may be scenarios where they match to some degree.

What would those particular scenarios be? A first situation in which the true individual-level likelihood may be different from the composite likelihood because datasets share biological individuals is when physiological trade-offs manifest within individuals, \eg individuals that have good reproduction one year have bad survival the next (a dynamic trade-off). This affects the probabilistic independence assumption between the survival and reproduction components of the model, which have to be specified with a joint probability. Because these terms are deep within eq.~\ref{surv-repro-indiv-LL}, it seems unlikely that a quick workaround to writing out the full likelihood of those models can be found, with trade-offs specified at the individual level. If one uses a composite likelihood in this case, one would be fitting essentially an approximation of the true model of unknown quality. A second situation which can generate an individual-level, full likelihood different from the composite likelihood pertains to having correlated individuals, \eg genetically related individuals with similar traits, or individuals that belongs to different clusters (\eg behavioural types). As this affects the very last simplification on the way to the component product likelihood, there is good hope to be able to reduce the complexity of the individual-level likelihood in this case (even analytically, using results for mathematically similar Hidden Markov Models, \citealt{zucchini2017hidden}). It is unclear whether using a  $m$-array formulation (see \eg \citealt{kery2011bayesian}) in such IPMs would be feasible. The two complexities mentioned above, dynamic trade-offs between reproduction and survival as well as correlated individuals, are often not included in IPM robustness checks that have evaluated the effects of (physical) dependence between datasets (even though those are rather hotly debated topics in ecology at large, \citealt{authier2017wolf}). It would be most interesting to evaluate the robustness of IPMs specified as composite likelihoods to physical dependence with either dynamic trade-offs or correlated individuals (see similar suggestions by \citealt{schaub2021integrated}). 

At this point I should stress that one can build IPMs without a probabilistic independence assumption, but rather with a conditional independence assumption. By this we mean that there is a sequential dependence of datasets, which often provides similar-looking composite likelihoods. The probabilistic independence assumption is only paramount if the composite likelihood is written as $\mathbb{P}(\mathbf{D}_1|\boldsymbol{\theta}_1) \mathbb{P}(\mathbf{D}_2|\boldsymbol{\theta}_2) \mathbb{P}(\mathbf{D}_3|\boldsymbol{\theta}_3)$. But there is no special restriction about this, \eg one could just as well write a model of the form $\mP(\mathbf{D}_1 \cap \mathbf{D}_2 \cap \mathbf{D}_3 |\boldsymbol{\theta}) = \mP(\mathbf{D}_3 | \mathbf{D}_1,\mathbf{D}_2,\boldsymbol{\theta}_3) \mP(\mathbf{D}_2 | \mathbf{D}_1,\boldsymbol{\theta}_2) \mP(\mathbf{D}_1|\boldsymbol{\theta}_1)$ as pointed out in \citet{frost2023integrated}. This is actually done for most models with density-dependence, as survival probabilities and reproduction parameters are then affected by the population count/densities provided by the demographic and count model (which would be $\mathbf{D}_1$ in the example above). Such non-independence often goes unrecognized because the Bayesian BUGS-style format in which many IPMs are specified allows to easily create such dependencies. But, importantly, such a model is \textit{still formulated through a composite likelihood built at population-level}, not individual-level. In a Bayesian setup with MCMC, we do not even need to use component likelihoods products---one could craft the desired dependencies at various hierarchical levels, including the individual one. Individual-level IPMs come with their own challenges though: while an individual-level model may be estimated for really small datasets, an individual-level likelihood with hundreds to thousands of individuals remains undesirable. Approximate Bayesian computation methods for individual-based models have been developed but remain challenging \citep{siren2018assessing}. 

Therefore, composite likelihoods are still very useful and a key question remains: how often are the individual-based (full) and composite likelihoods equal? Beyond the arguably toy-like models considered in the present paper, that is an open question for which we have yet no clear answer. \citet{beliveau2016data} made some progress on IPMs of a slightly higher complexity including demography, a population census, and a Jolly-Seber model for capture-recapture, by comparing numerical estimates obtained with the full likelihood to those obtained with the composite likelihood. Unfortunately, while these IPMs are more realistic depictions of what ecologists want to fit to data, they are also less mathematically tractable than the ones considered here. In \citet{beliveau2016data}, the $m$-array representation of the CMR data combined to counts led to independence of likelihood components only when conditioning on the latent states, which does not allow to separate the full likelihood into its component products. There were some small differences between analyses using full and composite likelihoods, such as narrower credible intervals with the full likelihood, but these did not seem to be affected by the percentage of overlap in individuals between datasets. Hence although \citet{beliveau2016data} did not find that full and composite likelihoods were equal, the numerical results did suggest that the two likelihoods must be relatively close. 

Many IPMs are written as a combination of a CJS capture-recapture model to estimate survival rates under imperfect detection, some fecundity data---whenever nests or dens can be surveyed---and a model producing counts, usually through a combination of a matrix population model and an observation model of such counts. The CJS and the matrix population model share some information as they usually have equal survival probabilities, for instance. However, the capture-recapture part of an IPM most often does not include parameters related to reproduction or recruitment, and therefore contribute only indirectly to the estimation of the rate of change in population size. For this reason, the use of the CJS model (or similarly designed mark-recovery models), which focus on survival estimation of a subset of marked individuals, is perhaps behind some of our difficulties to map the individual-based survival and detection estimation to the demography and the counting process at an individual level. By contrast, the Jolly-Seber (JS) model, as hinted above, can provide population growth rates and abundance in addition to demographic parameters such as survival and recruitment. It is therefore possible that some progress in clarifying the relationship between individual-based and composite likelihoods in IPMs could be achieved by adopting a JS representation of the capture-recapture data more often, in cases where a lot of the individuals are marked. In relatively closed populations with almost no immigration, specifying a fully individual-based model that can be sampled with a JS model requires a latent branching process representation for the demographic model, where individuals are born from their parents, rather than simply new recruits entering the population at various times (as in most formulations of the JS). This makes the demographic and capture-recapture components of the model more compatible, and should as a bonus provide some opportunities to bring closer statistical and theoretical ecology. 

\section*{Acknowledgements}

Exchanges with Olivier Gimenez and Matthieu Paquet motivated me to investigate further what independence meant in IPM papers, leading to a presentation at EURING 2023 of an earlier version of this work. Rachel McCrea \& Fay Frost presented on that occasion a removal-based model that also leads to proportionality between full and composite likelihood, and Charles Yackulic pointed out to me the relevance of Jolly-Seber capture-recapture models to IPMs. I then corresponded with Audrey Béliveau who also provided insights into her work on IPMs. I thank Matthieu Paquet and Christie le Coeur for comments on the manuscript. 

\bibliographystyle{amnat}
\bibliography{independence_IPMs}

\newpage

\subsection*{Appendix 1: Marginalized joint capture-recapture and reproduction model}

Since the capture-recapture-reproduction model of the main text can be written as a Hidden Markov Model (where $Z_{i,t} \in \{1,0\}$ is the latent state for individual $i$ at time $t$), it is tempting to try to use the formula for the likelihood of such an individual-based model using the Forward algorithm \citep{zucchini2017hidden,mcclintock2020uncovering}. For the CJS (capture-recapture) part of the model, for individual $i$, we have the likelihood: 

\begin{equation}\label{eq:LL-forward-algo}
    \mathcal{L}_{\text{capture}}(\boldsymbol{\varphi}; \mathbf{x}_i) = \boldsymbol{\delta} \mathbf{P}(x_1) \boldsymbol{\Gamma}_2  \mathbf{P}(x_2) \boldsymbol{\Gamma}_3  \mathbf{P}(x_3) ... \boldsymbol{\Gamma}_T \mathbf{P}(x_T) \mathbf{1}
\end{equation}

with the transition matrix for hidden states $$
    \boldsymbol{\Gamma}_t = 
    \begin{pmatrix} 
      \phi_{i,t} &  1-\phi_{i,t}\\
      0 & 1
\end{pmatrix}$$
and the observation probabilities conditional to the true states
$$
\mathbf{P}(x_t) = 
\begin{pmatrix} 
      p^{x_t} (1-p)^{1-x_t} & 0\\
      0 & 1-x_t
\end{pmatrix}.$$ 
In order to have reproduction as well in the model, we can replace $\mathbf{P}(x_t)$ by $\mathbf{P}(y_t,x_t) = \mathbf{P}(y_t|x_t)\mathbf{P}(x_t)$ using conditional probabilities and a diagonal $\mathbf{P}(y_t|x_t) =
\begin{pmatrix} 
      r_{\lambda}(y_t)^{x_t} & 0\\
      0 & 1
\end{pmatrix}$ representing reproduction measurement conditional to recapture. Unfortunately, while diagonal matrix multiplication is commutative (thus $\mathbf{P}(y_t|x_t)\mathbf{P}(x_t) = \mathbf{P}(x_t)\mathbf{P}(y_t|x_t)$), in general matrix multiplication is not commutative (e.g., of $\mathbf{P}(y_t|x_t)$ with $\boldsymbol{\Gamma}_t$) which does not allow to separate a variant of eq.~\ref{eq:LL-forward-algo} into two matrix subproducts, one for reproduction and one for survival. Tedious matrix multiplication of eq.~\ref{eq:LL-forward-algo} for $T=3$ leads to 
\begin{equation}\label{eq:LL-forward-algo-devel}
\begin{split}
    \mathcal{L}_{\text{capture}}(\boldsymbol{\varphi}; \mathbf{x}_i)  = \delta_1 p^{x_1} (1-p)^{1-x_1} 
    \left( \phi_2 p^{x_2} (1-p)^{1-x_2} \right.
    \left( \phi_3 p^{x_3} (1-p)^{1-x_3} + (1-\phi_3)(1-x_3) \right) \\
    \left. +(1-\phi_2) (1-x_2)(1-x_3) \right) + \delta_0 (1-x_1) (1-x_2) (1-x_3). 
\end{split}
\end{equation}
Applying the equation for $\mathbf{P}(x_1)=I$ and constant survival probabilities leads to the same results as \citet{yackulic2020need}. A version of eq.~\ref{eq:LL-forward-algo} including $\mathbf{P}(y_t|x_t)$ would lead to the scalar
\begin{equation}\label{eq:LL-forward-full}
\begin{split}
        \mathcal{L}_{\text{full}}(\boldsymbol{\varphi},\boldsymbol{\lambda}; \mathbf{x}_i,\mathbf{y}_i) = \delta_1 p^{x_1} (1-p)^{1-x_1} \mathbb{P}(Y_1=y_1|X_1 = x_1)
    \left( \phi_2 p^{x_2} (1-p)^{1-x_2} \mathbb{P}(Y_2=y_2|X_2 = x_2) \right.\\
    \left( \phi_3 p^{x_3} (1-p)^{1-x_3} \mathbb{P}(Y_3=y_3|X_3 = x_3) + (1-\phi_3)(1-x_3) \right) \\
    \left. +(1-\phi_2) (1-x_2)(1-x_3) \right) + \delta_0 (1-x_1) (1-x_2) (1-x_3). 
\end{split}
\end{equation}
which we can rewrite with $\mathbb{P}(Y_t = y_t|X_t = x_t) = r_{\lambda}(y_t)^{x_t}$. Matrix non-commutativity makes it impossible that reproduction and survival data can be separated in matrix product form, but such separation is possible in the scalar version
\begin{equation}\label{eq:LL-forward-full-simple}
\begin{split}
    \mathcal{L}_{\text{full}}(\boldsymbol{\varphi},\boldsymbol{\lambda}; \mathbf{x}_i,\mathbf{y}_i)  = \delta_1 p^{x_1} (1-p)^{1-x_1} r_{\lambda}(y_1)^{x_1}
    \left( \phi_2 p^{x_2} (1-p)^{1-x_2} r_{\lambda}(y_2)^{x_2} \right.\\
    \left( \phi_3 p^{x_3} (1-p)^{1-x_3} r_{\lambda}(y_3)^{x_3} + (1-\phi_3)(1-x_3) \right) \\
    \left. +(1-\phi_2) (1-x_2)(1-x_3) \right) + \delta_0 (1-x_1) (1-x_2) (1-x_3). 
\end{split}
\end{equation}

Let us consider the various possible cases, with constant $\phi$, $p$ and $\lambda$ over time. At time $t=1$, we assume no reproduction as well as certain capture for comparison to \citealt{yackulic2020need}. The assumptions yield: 
\begin{equation}\label{eq:LL-forward-full-111}
\begin{split}
    \mathcal{L}_{\text{full}}\left(\boldsymbol{\varphi},\boldsymbol{\lambda}; (1,1,1),(0, y_2,y_3)\right)  = \phi p r_{\lambda}(y_2) (\phi p r_{\lambda}(y_3)) = \phi^2 p² \times r_{\lambda}(y_2) r_{\lambda}(y_3)
\end{split}
\end{equation}
\begin{equation}\label{eq:LL-forward-full-110}
\begin{split}
    \mathcal{L}_{\text{full}}\left(\boldsymbol{\varphi},\boldsymbol{\lambda}; (1,1,0),(0, y_2,0)\right)  = \phi p r_{\lambda}(y_2) (\phi (1-p)+1-\phi) = (\phi^2 p (1-p)+(1-\phi)\phi p) \times r_{\lambda}(y_2)
\end{split}
\end{equation}
\begin{equation}\label{eq:LL-forward-full-101}
\begin{split}
    \mathcal{L}_{\text{full}}\left(\boldsymbol{\varphi},\boldsymbol{\lambda}; (1,0,1),(0,0,y_3)\right)  = \phi (1-p) (\phi p r_{\lambda}(y_3)) = \phi^2 p(1-p) \times r_{\lambda}(y_3)
\end{split}
\end{equation}
\begin{equation}\label{eq:LL-forward-full-100}
\begin{split}
    \mathcal{L}_{\text{full}}\left(\boldsymbol{\varphi},\boldsymbol{\lambda}; (1,0,0),(0,0,0)\right)  = \phi (1-p) (\phi (1-p)+1-\phi) + (1-\phi). 
\end{split}
\end{equation}
In all four cases the reproduction part of the likelihood can be factorized out. It might be provable by recurrence that it is always the case, since eq.~\ref{eq:LL-forward-full-simple} can be extended to $T$ time steps and has always the same structure. 

\end{document}